\documentclass{aa}  
\usepackage{graphicx}
\usepackage{txfonts}
\usepackage{natbib,twoopt}
\usepackage{soul}
\bibpunct{(}{)}{;}{a}{}{,}             %% natbib format for A&A and ApJ
\usepackage{hyperref}
\usepackage[draft]{changes}
\hypersetup{
    colorlinks=true,
    linkcolor=blue,
    citecolor = blue,
    filecolor=magenta,      
    urlcolor=blue}

\definechangesauthor[name={AM}, color=red]{AM}
\graphicspath{{./}{figures/}}

\begin{document} 

\authorrunning{Alvarez Garay et al.}
\titlerunning{MgAl burning chain in M~54}

\title{MgAl burning chain in M~54: \\
the globular cluster-like properties of a nuclear star cluster 
}

\author{D. A. Alvarez Garay\inst{1},
    A. Mucciarelli\inst{2}\fnmsep\inst{3}, 
    P. Ventura\inst{4}, 
    M. Bellazzini\inst{3},
    and S. Covella\inst{2}\fnmsep\inst{3}
}

   \institute{Osservatorio Astrofisico di Arcetri, Largo E. Fermi 5, 50125
              Firenze, Italy\\
              \email{deimer.alvarez@inaf.it}
        \and
        Dipartimento di Fisica e Astronomia, Università degli Studi di Bologna, Via Gobetti 93/2, I-40129 Bologna, Italy
        \and 
       INAF, Osservatorio di Astrofisica e Scienza dello Spazio di Bologna, Via Gobetti 93/3, I-40129 Bologna, Italy
       \and
       INAF, Osservatorio Astronomico di Roma, Via Frascati 33, 00078, Monteporzio Catone, Rome, Italy
}

\abstract
{In this study, we present the chemical abundances of Fe, Mg, Al, Si, and K for a sample of 233 likely member stars of M~54. All the stars were observed with the FLAMES high-resolution multi-object spectrograph mounted at the VLT.  Our analysis confirmed the presence of a large metallicity range in M~54, with the majority of the stars having $\rm -1.8 <[Fe/H]< -1.0$ dex and few stars with [Fe/H] $> -1.0$ dex. The mean value of the total sample is $\rm [Fe/H] = -1.40$ ($\sigma = 0.22$ dex). A Markov Chain Monte Carlo analysis revealed that the observed spread in [Fe/H] is compatible with a non-null intrinsic iron dispersion.  We also found that the metallicity distribution function and the broadening of the red giant branch of M~54 are not compatible with a single age, but instead they suggest a wide age range from $\sim13$ Gyr to $\rm \sim1-2$ Gyr or a smaller age range if a significant He enhancement ($\rm Y \sim 0.35/0.40$) is present in the most metal-rich stars.
We identified among the stars in M~54 the entire pattern of anticorrelations linked to the MgAl burning cycle. In particular, the metal-rich component displays a higher level of H-burning with the presence of more extended anticorrelations than the metal-poor component. No Mg-poor ([Mg/Fe]$<$0.0 dex) 
stars are identified in M~54.
The evidence collected so far cannot be explained neither with a globular cluster-like scenario nor with a galactic chemical evolution. 
The chemical properties of M~54 can be explained within a scenario where this system formed through the merging of two globular clusters, the metal-poor one with standard characteristics and the more metal-rich one with more pronounced chemical anomalies, a possibly younger than the first one. 
M~54 is confirmed as a key stellar system for explaining the chemical evolution of a nuclear star cluster.
}

\maketitle
%
%-------------------------------------------------------------------
\section{Introduction} \label{intro}

Nuclear star clusters (NSCs) are among the densest stellar systems present in the Universe (\citealt{walcher_05,norris_14}), with masses of $\sim 10^6$ up to $\rm 10^8 ~M_{\odot}$ and half-light radii of about 1 to 10 pc (\citealt{georgiev_14,georgiev_16}). 
They exhibit structural and stellar-population complexities that point to extended evolutionary histories \citep[see e.g.][]{neumayer}, with age and metallicity spreads, and evidence for prolonged or recurrent star formation (\citealt{walcher_05,kacharov_18}).
NSCs form through a combination of mechanisms: in-situ star formation fueled by gas inflows toward the galactic nucleus, and inspiral and merging of star clusters driven by dynamical friction, with this latter channel preferred in galaxies with $\rm M_{\star}< 10^9~M_{\odot}$ \citep{neumayer}. 
Determining the relative importance of these channels is essential for understanding the co-evolution of NSCs and their host galaxies.

The intrinsically compact nature of NSCs, together with the large distances to most galaxies that host them, generally prevents resolving them into individual stars. In this context, two nearby systems offer unique and powerful laboratories: $\omega$ Centauri and M~54. 
They are usually classified as globular clusters (GCs) according to their morphology but exhibits a higher level of complexity (akin to NSCs) with respect to the majority of GCs, for instance intrinsic [Fe/H] spread. Nevertheless, they display some chemical patterns (i.e. Na-O and Mg-Al anticorrelations) that are considered distinguishing features of GCs.
Both objects are close enough to enable high-precision photometry and spectroscopy of their individual stars, allowing us to examine a NSC (or the relic of one) with a level of detail unattainable for any extragalactic counterpart. This makes them fundamental benchmarks for testing theoretical models of NSC formation and for linking the integrated-light properties of unresolved NSCs to the underlying stellar populations.

While $\omega$ Centauri is considered as the possible remnant of a disrupted dwarf spheroidal galaxy \citep{bekki03,bekki_19}, the association of M~54 with an external galaxy is direct: it is embedded at the center of the remnant of the Sagittarius dwarf spheroidal galaxy (Sgr dSph).
Sgr dSph is a perfect example of a Galactic satellite that is currently undergoing disruption by the tidal field of the Milky Way (\citealt{ibata_94,ibata_20,majewski_03,ramos_22}). 
The central region of Sgr, often referred as the NSC of the galaxy, exhibits a composite stellar population in terms of ages and metallicities, with a bimodal [Fe/H] distribution \citep{carretta_10,mucciarelli_17,alfaro_cuello_19}, with the two main peaks associated to the red giant branches (RGBs) visible in the color-magnitude diagram (CMD) of Sgr.
The metal-poor peak of the [Fe/H] distribution, at [Fe/H]$\sim$--1.5/--1.6 dex, is dominated by the old, GC-like system M~54, while the second peak, at [Fe/H]$\sim$--0.5 dex, corresponds to a Sgr population with ages of $\sim$4-6 Gyr.
There is now some consensus on a scenario in which the stellar nucleus of the Sgr dSph galaxy was formed by the concurrence of in situ star formation and the infall of one or more GCs by dynamical friction \citep{bellazzini_08,carretta_10,alfaro_cuello_19,alfaro_cuello_20}. In this paper we refer to M~54 as the bulk of the old and metal-poor population of the Sgr NSC, the one that has been considered as a MW GC until the discovery of the Sgr dSph galaxy, following \citet{bellazzini_08}.

The chemical properties of M~54 further reinforce its somehow dual nature. Its stars exhibit a pronounced spread in [Fe/H] \citep{bellazzini_08,carretta_10,mucciarelli_17,alfaro_cuello_19}, implying a sufficiently deep potential to retain supernova (SN) ejecta and sustain extended chemical evolution, behavior not observed in Milky Way GCs but typical of galactic systems. 
At the same time, M~54 shows the classic light-element anticorrelations (e.g. Na–O, Mg–Al) characteristic of multiple populations (MPs) in massive GCs \citep{carretta_10,fernandez_21}. 
The chemical anticorrelations observed in all GCs are interpreted as the result of self-enrichment within the cluster \citep[see e.g.,][]{bastian_18}, 
where low-velocity material processed through the hot CNO cycle and its secondary NeNa and MgAl chains (e.g., \citealt{langer_93}; \citealt{prantzos_07}) is incorporated in a subsequent generation of stars. 
The most accepted theoretical models for the formation of MPs involve the occurrence of at least two episodes of star formation where a second generation (SG) of stars, enriched/depleted in the elements of the CNO cycle formed from the material polluted by the first generation (FG) or polluter stars within the first 100-200 Myr of the cluster life. 
Different polluters were proposed in the literature, including intermediate- or high-mass stars in their asymptotic giant branch (AGB) phase \citep{dercole_10}, fast rotating massive stars  \citep{krause_13}, novae \citep{maccarone_12,denissenkov_14}, interacting binary stars \citep{mink_09}, and supermassive stars \citep{denissenkov_hart_14}. All these polluters suffer in the complete explanation of the observed evidence. 

The coexistence of a broad iron dispersion together with GC-like light-element patterns suggests that M~54 hosts processes commonly associated both with NSCs and with dense cluster formation environments, highlighting its dual nature.
These properties make M~54 uniquely valuable for probing the formation pathways of NSCs and for bridging the gap between resolved stellar populations in the Local Group and the unresolved NSCs observed in more distant galaxies. 

In this work we present a detailed chemical analysis for 233 likely members of M~54, observed with the high-resolution spectrograph FLAMES@VLT. 
This paper is organized as follows: in Sect.~\ref{obs} we present the data, in Sect.~\ref{params} we describe the stellar parameters and velocities, in Sect.~\ref{abu} we detail the analysis, in Sect.~\ref{results} we illustrate the results of the abundance analysis, in Sect.~\ref{discussion} we carry out the discussion of the results, and in Sect.~\ref{summary} we summarize our work.

%-------------------------------------------------------------------
\section{Observations and target selection} \label{obs}
The employed spectra were acquired at the Very Large Telescope UT2 (Kueyen) with the optical multi-object spectrograph FLAMES \citep{pasquini_02} under the programmes 075.D-0075 (P.I. Mackey, from July to August 2005), 081.D-0286 (P.I. Carretta, from June to September 2009), and 095.D-0539 (P.I. Mucciarelli, from July to August 2015). 
FLAMES/GIRAFFE was employed in the high resolution mode that allows to allocate up to 132 fibers simultaneously, while FLAMES/UVES allows to allocate eight high resolution fibers. 

Within the programme 075.D-0075 was adopted the GIRAFFE HR21 setup (R=18000 and a wavelength coverage $\sim$ 8484-9000 $\AA$). 
Under the programme 081.D-0286 were employed two GIRAFFE setups: the HR11 (R=29500 and a wavelength coverage $\sim$ 5597-5840 $\AA$) and the HR13 (R=26400 and a wavelength coverage $\sim$ 6120-6405 $\AA$). 
Finally, within the programme 095.D-0539 was used the GIRAFFE+UVES combined mode. The adopted setups are the HR18 (R=20150 and a wavelength coverage $\sim$ 7468-7889 $\AA$), and the UVES Red Arm 580 (R=45000 and wavelength coverage $\sim$ 4800-6800 $\AA$).
All the used setups allow us to measure Fe, Mg, Al, Si, and K abundances. 
All the spectra were reduced using the dedicated GIRAFFE and FLAMES/UVES ESO pipelines\footnote{\url{https://www.eso.org/sci/software/pipelines/giraffe/giraffe-pipe-recipes.html}\\ \url{https://www.eso.org/sci/software/pipelines/uves/}}, which include the bias subtraction, flat field correction, spectral extraction, wavelength calibration and (for the UVES fibers) order merging.

\begin{figure}[h]
\centering
\includegraphics[width=9 cm]{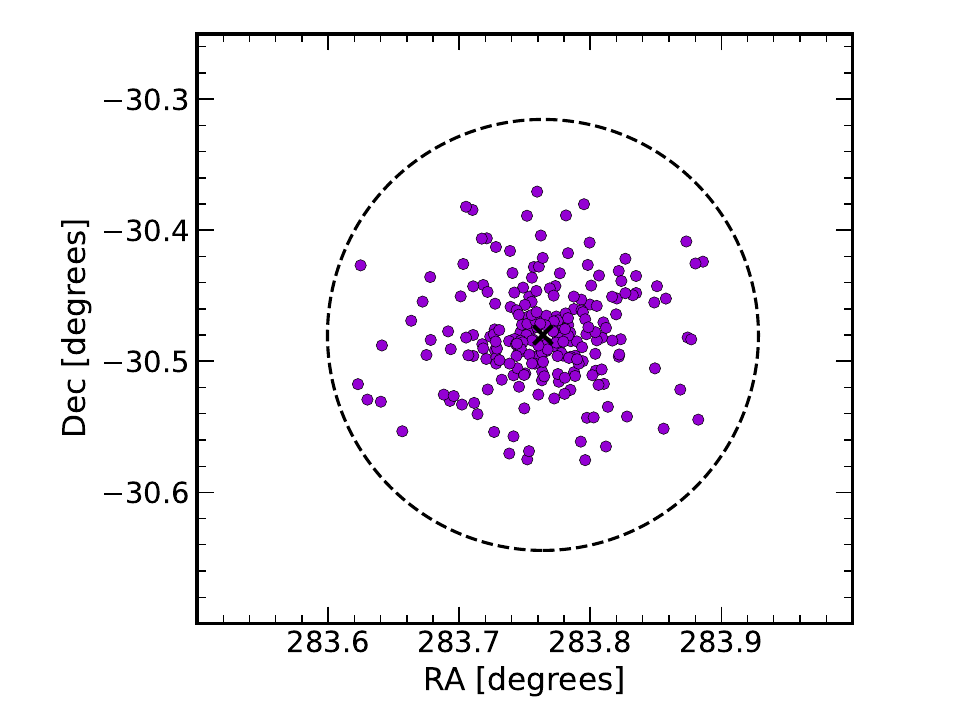}
\caption{Coordinate positions of the observed targets are displayed by the purple circles. The black cross denotes the cluster center \citep{baumgardt_18}. 
The dashed black circle displays the tidal radius \citep{harris_10}.}
\label{fig:spa_dist}
\end{figure}

\begin{figure}[h]
\centering
\includegraphics[width=9 cm]{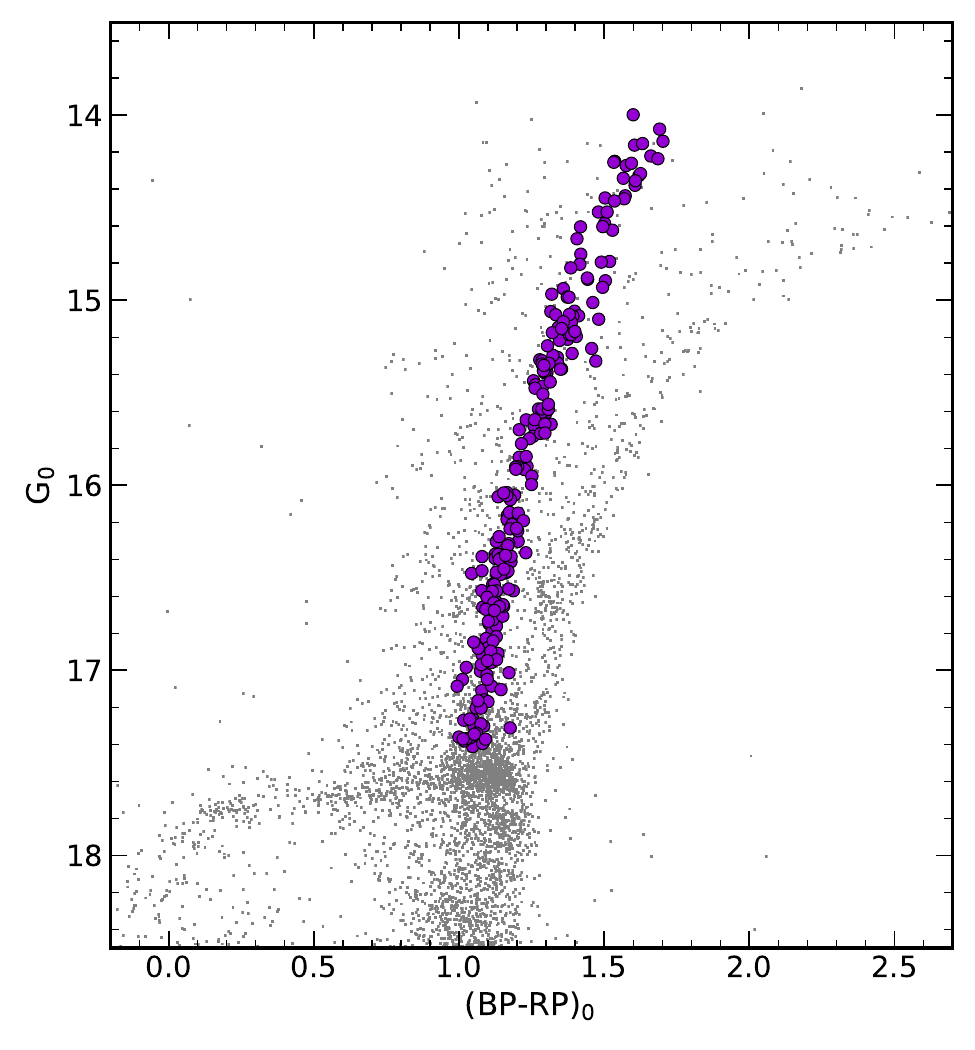}
\caption{CMD of M~54. Gray points represent the targets associated to Sgr according to the proper motions of Gaia DR3, 
while the purple circles represent the spectroscopic target stars.} \label{fig:cmd}
\end{figure}

All the used programs sample the spatial region within the tidal radius of M~54, namely $r_t = 9\rlap{.}^{'}868$ \citep{harris_10}.
We considered only stars located along the RGB of M~54 (the bluest portion of the RGB of the nuclear region of Sgr), excluding stars belonging to the reddest RGBs (therefore stars younger and/or more metal-rich). 
The bluest RGB is dominated by M~54 with superimposed the old, metal-poor Sgr population \citep[see][]{minelli23,liberatori25}.
We discarded stars with signal-to-noise ratio (SNR) $<$20 (mainly stars with G$>$17) because these spectra do not provide reliable abundances. 
According to their measured radial velocities (RV, see Sect.~\ref{rv}) and their velocity dispersion profiles we identified as likely members of M~54 those stars with 120 km s$^{-1} \lesssim$ RV $\lesssim$ 170 km s$^{-1}$, 
following the procedures described in \citet{bellazzini_08}. At the end, the sample analyzed in this work includes 233 stars. 
In Table~\ref{tab:1} are reported the total number of stars analyzed for the different combination of setups.  In particular, we analyzed a total of 211 stars from the programme 075.D-0075, 42 stars from the programme 081.D-0286, and 87 stars from the programme 095.D-0539. As we can see from Table~\ref{tab:1} many stars are in common among the three programmes. 
Figures \ref{fig:spa_dist} and \ref{fig:cmd} show the spatial distribution and the position in the Gaia CMD of the entire sample, respectively.

We inspected  the RUWE parameter given by the \textit{Gaia} catalog for all the targets. According to the \textit{Gaia} documentation the RUWE is expected to be around 1.0 for sources where the single-star model provides a good fit to the astrometric observations. The generally adopted threshold value that could indicate that the source is non-single or otherwise problematic for the astrometric solution is 1.4 \citep[][and references therein]{gaia_23}. In our sample, there are a total of 25 stars with RUWE $> 1.4$ and among them there are nine stars with $2.0<$ RUWE $<4.0$ . However, we decided to maintain these stars in the dataset as they do not display any significant anomaly neither in their RVs nor in their abundances. 

\begin{table}[h]
\caption{Stars observed for different combination of setups.}
\label{tab:1}
\centering
\renewcommand{\arraystretch}{1.5}
\scriptsize
\setlength{\tabcolsep}{9pt}
\begin{tabular}{cc}
\hline\hline 
Setups & N \\
\hline
HR11, HR13, HR18, HR21 & 21  \\
HR11, HR13, HR18 & 11 \\
HR13, HR18, HR21 & 7 \\
HR13, HR18 & 3 \\
HR18, HR21 & 32 \\
HR18 & 8 \\
U580, HR21 & 5 \\
HR21 & 146 \\
\hline 
TOTAL & 233 \\
\hline\hline
\end{tabular}
\vspace{2pt}
\end{table}

The stars observed with the setups HR11 and HR13 were already analyzed by \citet{carretta_10}, while the stars observed with the setup HR18 and in common with \citet{carretta_10} were analyzed by \citet{carretta_22}. Also, a total of 83 stars in our dataset are present among the stars analyzed by \citet{bellazzini_08} with the  multi-object spectrographs DEIMOS on the Keck 2 telescope and with the HR21 setup, and among them 37 stars are in common also with the dataset analyzed by \citet{carretta_10}. All the other stars present in our sample were never analyzed before making this one the largest database of chemical abundances of M~54 stars based on high-resolution spectra. 

%--------------------
%--------------------
%--------------------

%-------------------------------------------------------------------
\section{Atmospheric parameters and radial velocities} \label{params}
%--------------------
\subsection{Atmospheric parameters} \label{atm_params}
The atmospheric parameters were derived by using  the photometric information from the \textit{Gaia} Data Release 3 \citep{gaia_16,gaia_23}, 
in order to avoid the spurious effects introduced by the spectroscopic determination of the parameters \citep{mb20}.
We obtained the dereddened $\rm (BP-RP)_0$ color by assuming a color excess factor $\rm E(B-V) = 0.15\pm0.03$ \citep{harris_10} and adopting the iterative recipe proposed by \citet{gaia_18}.
We derived the effective temperatures ($\rm T_{eff}$) for all the targets from the empirical $\rm (BP-RP)_0-T_{eff}$ relation by \citet{mucciarelli_21}, calibrated on stars whose temperature was obtained with the infrared flux method. Gaia BP/RP photometry, due to the relatively large apertures adopted, can suffer from contamination in crowded fields \citet[][and references therein]{riello_21}. To verify if this affects the temperatures inferred from $\rm (BP-RP)_0$ colors we used the $\rm C^*$ parameter, introduced by \citet{riello_21} as an indicator of the degree of contamination in the BP and RP aperture windows compared to the smaller window used for the G photometry. For 80 stars in our sample with very high $\rm C^*$ values, $\rm C^* > 3$, we got $\rm (V-I)$ colors from the catalog by \citet{monaco_02}, obtained from PSF fitting photometry. We used the transformations present in the \textit{Gaia} documentation\footnote{\url{https://gea.esac.esa.int/archive/documentation/GDR3/Data_processing/chap_cu5pho/cu5pho_sec_photSystem/cu5pho_ssec_photRelations.html}} to transform the $\rm (V-I)$ in $\rm (BP-RP)$ color, then we applied the reddening corrections and the \citet{mucciarelli_21} relations as above, to obtain independent estimates of $\rm T_{eff}$ for these eighty stars.

The average differences between the temperatures directly derived from \textit{Gaia} photometry and those obtained from the transformed magnitudes of \citet{monaco_02} are $\sim 30$ K, which translates in differences in the derived abundances of $\sim0.02-0.03$ dex. We conclude that contamination in the crowded environment of M~54 does not affect $\rm (BP-RP)$ colors and, consequently, our inference of $\rm T_{eff}$ for the target stars in any significant way, for our purpose.

We calculated the internal errors in $\rm T_{eff}$ as the sum in quadrature of the errors due to the uncertainties in photometric data, reddening, and $\rm (BP-RP)_0-T_{eff}$ relation. The errors are of the order of $\sim 90-130$ K. 

To validate our photometric temperatures, we performed a sanity check on the subset of stars with a high number of available Fe I lines (specifically those observed with UVES and the HR11, HR13, HR18, and HR21 setups). For these targets, we derived temperatures via the excitation equilibrium method, which assumes no correlation between iron abundance A(Fe) and excitation potential $\chi$. The spectroscopic temperatures show excellent agreement with the photometric values, with differences typically $\rm \leqslant 50~K$. This corresponds to a negligible impact on derived abundances ($\sim 0.04-0.05$ dex or less), confirming the reliability of our photometric temperatures for the entire sample.

Regarding the surface gravities ($\rm \log g$) we adopted the Stefan-Boltzmann relation, using the photometric temperature described above and assuming a stellar mass of $\rm 0.80~M_{\odot}$, 
according to a BaSTI isochrone with age 13 Gyr and [Fe/H]$\sim$--1.5 dex \citep{hidalgo_18}. Stellar luminosities were computed by using the dereddened G-band magnitude, 
the bolometric corrections by \citet{andrae_18} and a true distance modulus $\rm (m-M)_0 = 17.10\pm0.15$ mag \citep{monaco_04}. 
By propagating the uncertainties in $\rm T_{eff}$, distance modulus, and photometry we obtained uncertainties in $\rm \log g$ of the order of $\sim 0.1$.

We derived the microturbolent velocities ($\rm v_t$) from the $\rm \log g - v_t$ relation provided by \citet{kirby_09}, assuming an error of 0.2 km s$^{-1}$.  
We did not determine $\rm v_t$ values spectroscopically, in order to avoid large fluctuations, potentially caused by the small number of Fe~I lines available for most of the targets. 

The derived atmospheric parameters, together with additional information are reported in Table~\ref{tab:2}.

%--------------------
\subsection{Radial velocities} \label{rv}
We determined RVs by using the standard cross-correlation technique implemented in the \texttt{IRAF} task \texttt{FXCOR}. Synthetic spectra generated with the \texttt{SYNTHE} code \citep{sbordone_04,kurucz_05} were employed as template spectra, convolved with a Gaussian profile to reproduce the observed spectral resolution.

When more than one spectrum per star was available, the final heliocentric RV was calculated as the mean of the individual RV values. Table \ref{tab:2} presents the final heliocentric RVs for all targets. The uncertainties reported are calculated as the dispersion of the mean RV normalized to the root mean square of the number of used exposures; 
when only one spectrum per star was available, we used the error provided by \texttt{FXCOR}. 

We checked for possible RV variations among different spectra: only one star, namely \#3801447, shows signs of RV variations, with differences of $\rm 3~km~s^{-1}$ and a RUWE $= 3.73$. For all the other stars (including those with RUWE$>$1.4) the RVs are consistent within the uncertainties. 

\begin{figure}[h]
\centering
\includegraphics[width=9 cm]{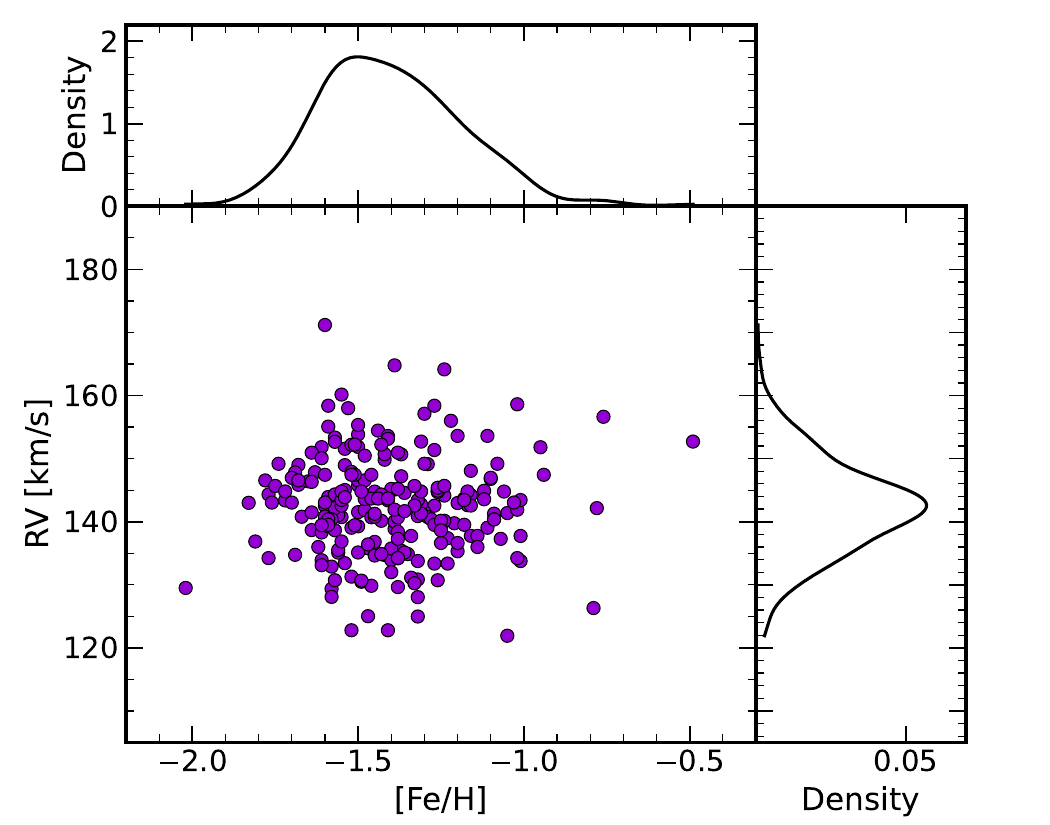}
\caption{Distributions of the [Fe/H] and RVs for the target stars. The main panel shows the behavior of the RV of the observed stars as a function of [Fe/H]. The generalized histograms of [Fe/H] and RV distributions are also plotted.}
\label{fig:rvfe}
\end{figure}

From the analysis of 233 stars we derived a final RV mean value of +142.3$\pm$0.5 ($\sigma = 7.8$) km s$^{-1}$. This value is perfectly in agreement with the mean values reported by \citet{carretta_10} of +143.7$\pm$0.9 ($\sigma = 8.3$) km s$^{-1}$, and by \citet{bellazzini_08} of +140.9$\pm$0.4 ($\sigma = 9.3$) km s$^{-1}$. 
Figure~\ref{fig:rvfe} shows the heliocentric RVs of the entire sample as a function of [Fe/H] and 
the RV distribution represented as a generalized histogram.
%--------------------
%--------------------

%-------------------------------------------------------------------
\section{Abundance analysis} \label{abu}
In this work we measured chemical abundances by using our own code \texttt{SALVADOR} (Alvarez Garay et al., in prep.), which performs a $\chi^2$ minimization between the line under analysis and a grid of synthetic spectra calculated with the appropriate atmospheric parameters and varying only the abundance of the considered element. We calculated the grids of synthetic spectra by using the \texttt{SYNTHE} code \citep{sbordone_04,kurucz_05}, and by computing one-dimensional, plane-parallel, local thermodynamic equilibrium (LTE) model atmospheres employing the \texttt{ATLAS9} code \citep{sbordone_04,kurucz_05} and 
the new opacity distribution functions of the KOALA database\footnote{\url{https://sites.google.com/view/koala-database/}} \citep{koala}. 

For abundance determination, we performed a thorough linelist selection across the spectral range covered by the used setups, selecting lines that are unblended, unsaturated, and not contaminated by telluric features at the resolution of selected setups, respectively. The atomic information for these lines was sourced from the Kurucz-Castelli linelist database\footnote{\url{https://wwwuser.oats.inaf.it/castelli/linelists.html}}, with some additional updates including the most recent laboratory measurements of $\log gf$ available in the literature. For the determination of the abundance ratios we adopted the solar reference from \citet{grevesse_98}. 

To determine star-to-star uncertainties associated with the chemical abundances we used the same approach described in \citet{alvarez_22}. The errors associated with the adopted atmospheric parameters were propagated by recalculating chemical abundances, varying only one parameter at a time ($\rm T_{eff}$, $\rm \log g$ or $\rm v_t$) by its uncertainty and keeping the other parameters fixed to their best value. Internal errors, associated with the measurement process, were estimated as the line-to-line scatter divided by the root mean square of the number of lines. When only one line was available, such in the case of K, the uncertainties were estimated by resorting to a Monte Carlo simulation. We created synthetic spectra with representative values for the atmospheric parameters of the analyzed stars, and we injected Poissonian noise, according to the SNR of the observed spectra. For each line, we created a total of 200 noisy spectra and derived the abundance with the same procedure used for observed spectra. Finally, we calculated the internal error as the standard deviation of the abundances derived from the 200 simulations.
\begin{table*}[h]
\caption{Data for the target stars belonging to M~54. The adopted solar abundances for the measured chemical elements are from \citet{grevesse_98} and they are reported in the header of each element abundance column.}
\label{tab:2}
\centering
\renewcommand{\arraystretch}{1.5}
\scriptsize
\setlength{\tabcolsep}{2.6pt}
\begin{tabular}{cccccccccccc}
\hline\hline 
ID & ID \textit{Gaia} & G & RV & T$_{\text{eff}}$ & $\log g$ & $v_t$ & [Fe/H] & [Mg/Fe] & [Al/Fe] & [Si/Fe] & [K/Fe] \\
 & {\it Gaia} DR3  & [mag] & [km s$^{-1}$] & [K] & [cgs] & [km s$^{-1}$] & 7.50 & 7.58 & 6.47 & 7.55 & 5.12 \\
\hline
M54\_1500138 & 6760429401842331136 & 14.5856 & $133.43\pm0.17$ & 4248 & 0.77 & 1.95 & $-1.54\pm0.11$ & $0.30\pm 0.03$ & $0.74\pm 0.05$ & $0.26\pm 0.13$ & $0.28\pm 0.15$ \\   
M54\_1500190 & 6760429157003643904 & 14.7517 & $134.73\pm0.38$ & 4332 & 0.89 & 1.93 & $-1.42\pm0.12$ & $0.34\pm 0.03$ & $0.55\pm 0.07$ & $0.21\pm 0.13$ & $0.21\pm 0.12$ \\  
M54\_1500360 & 6760449875922016384 & 15.2880 & $142.62\pm0.24$ & 4408 & 1.15 & 1.87 & $-1.17\pm0.12$ & $0.13\pm 0.03$ & $1.30\pm 0.05$ & $0.37\pm 0.14$ & $    ...     $ \\   
M54\_2300345 & 6760424831996623616 & 16.3033 & $150.67\pm0.20$ & 4712 & 1.72 & 1.73 & $-1.37\pm0.16$ & $0.47\pm 0.09$ & $1.05\pm 0.12$ & $    ...     $ & $0.49\pm 0.13$ \\  
M54\_2300384 & 6760413046606337920 & 16.3751 & $138.84\pm0.45$ & 4821 & 1.80 & 1.72 & $-1.25\pm0.16$ & $0.40\pm 0.09$ & $    ...     $ & $    ...     $ & $0.52\pm 0.11$ \\   
M54\_2300406 & 6760413080966064000 & 16.3376 & $141.51\pm0.20$ & 4798 & 1.78 & 1.72 & $-1.50\pm0.16$ & $0.12\pm 0.09$ & $    ...     $ & $    ...     $ & $0.51\pm 0.09$ \\   
M54\_2300585 & 6760449807202520960 & 16.7071 & $135.31\pm0.20$ & 4808 & 1.93 & 1.69 & $-1.20\pm0.16$ & $    ...     $ & $    ...     $ & $    ...     $ & $0.28\pm 0.09$ \\  
M54\_2406717 & 6760423079650001408 & 16.3158 & $135.39\pm0.48$ & 4766 & 1.75 & 1.73 & $-1.37\pm0.16$ & $0.36\pm0.06$ & $    ...     $ &  $    ...     $ & $0.44\pm0.10$ \\
M54\_2407142 & 6760423801205120256 & 14.4363 & $141.84\pm0.25$ & 4157 & 0.65 & 1.98 & $-1.48\pm0.09$ & $0.37\pm0.03$ & $0.01\pm0.05$ & $0.24\pm0.12$ & $0.09\pm0.15$ \\
M54\_2407725 & 6760424415355641728 & 17.1593 & $141.37\pm0.20$ & 4940 & 2.17 & 1.63 & $-1.05\pm0.19$ & $    ...     $ & $    ...     $ & $    ...     $ & $    ...     $ \\
\hline 
\end{tabular}
\tablefoot{This is a portion of the entire table, which is available in its entirety in at the CDS.}
\vspace{2pt}
\end{table*}
%%

%-------------------------------------------------------------------
\section{Results} \label{results}
In the following subsections, we will present in detail the results of the chemical analysis. 
The derived chemical abundance ratios are reported in Table~\ref{tab:2}, together with their final uncertainties. 
%--------------------
\subsection{Iron distribution} \label{iron_dist}

Figure~\ref{fig:rvfe} shows the metallicity distribution function (MDF hereafter) for the entire sample. 
The vast majority of the stars has [Fe/H] between --1.8 and --1.0 dex, with only one star with [Fe/H]$\sim$--2.0 dex and six stars with [Fe/H]$>$--1.0 dex (the latter is the boundary proposed by \citet{carretta_10} to discriminate between M~54 and Sgr stars).  
The total sample provides an average [Fe/H] of --1.40$\pm$0.01 ($\sigma$=~0.22 dex), while excluding the six stars with [Fe/H]$>$--1.0 dex and the most metal-poor one, the average value is --1.41$\pm$0.01 ($\sigma$=~0.19 dex). In both cases, the observed dispersion is significantly larger than the typical uncertainties in individual stars, indicating the presence of an intrinsic star-to-star scatter in [Fe/H] among the stars of 
the blue RGB. 
To confirm this we used a Markov Chain Monte Carlo (MCMC) approach based on the maximum-likelihood principle to assess whether the observed dispersion is compatible with a non-null intrinsic spread in [Fe/H], or whether it can be fully accounted for by the measurement errors. 
We ran the MCMC analysis both on the full sample of 233 stars and on the subsample of 226 stars with $\rm -2.0 < [Fe/H] < -1.0$ dex. For the complete sample, we obtained a mean value $\rm [Fe/H] = -1.41\pm0.01$ dex and an intrinsic scatter of $\rm \sigma_{int}=0.13 \pm 0.02$ dex, while for the subsample we derived an average $\rm [Fe/H] = -1.43\pm0.01$ dex and $\rm \sigma_{int}=0.09 \pm 0.02$ dex. 
These results indicate that the observed spread in [Fe/H] cannot be fully explained by observational uncertainties but instead 
it is compatible with a non-null intrinsic dispersion.
The average [Fe/H] that we found is in good agreement with \citet{bellazzini_08} ($\left\langle \text{[Fe/H]} \right\rangle = -1.45$ dex), while is $\sim 0.15$ and $\sim 0.10$ dex higher than the values found by \citet{carretta_10} and \citet{mucciarelli_17}, respectively. 

\begin{figure}[h]
\centering
\includegraphics[width=9 cm]{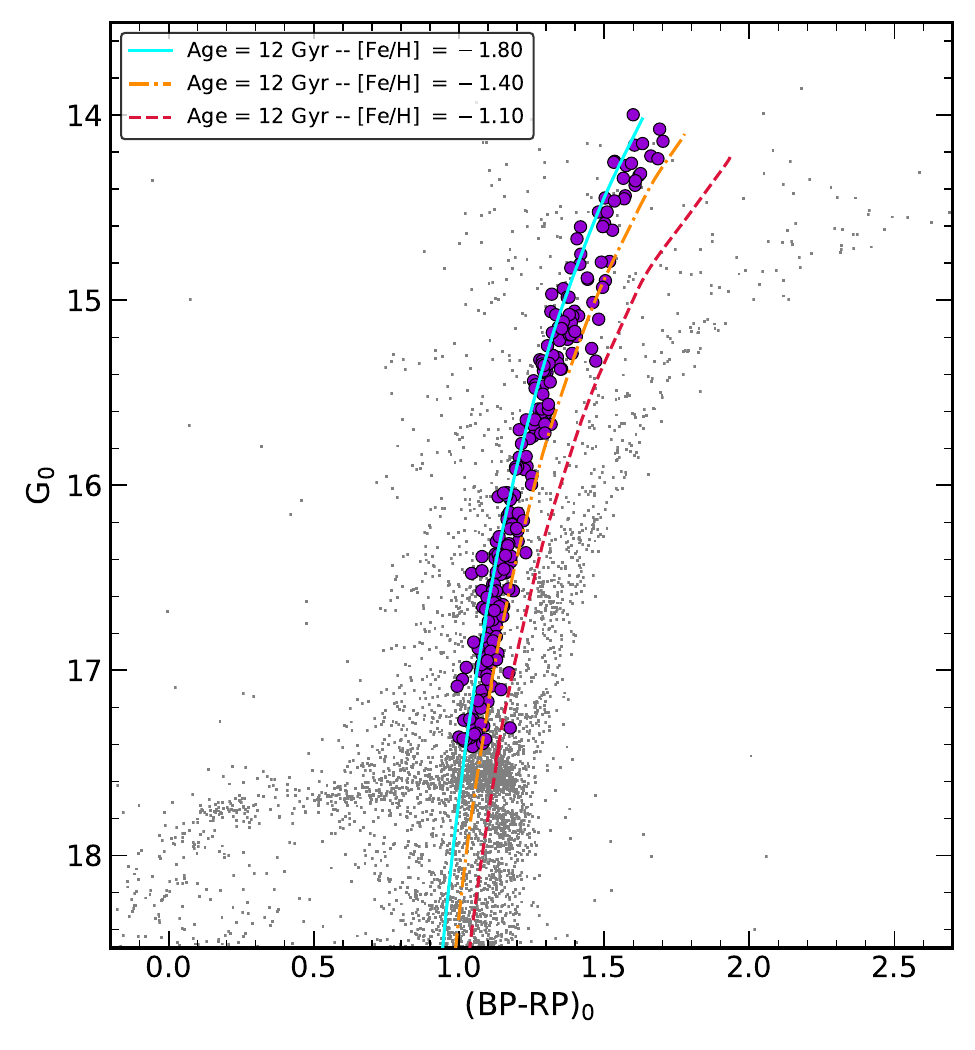}
\caption{CMD of M~54 with superimposed three different isochrones with the same age (12 Gyr) but a different [Fe/H]. The solid cyan isochrone has $\rm [Fe/H] = -1.8$ dex, the dash-dotted orange isochrone has $\rm [Fe/H] = -1.4$ dex, and the dashed red isochrone has $\rm [Fe/H] = -1.1$ dex.} \label{fig:cmd_iso}
\end{figure}

Despite M~54 is considered as an old stellar system \citep{siegel_07}, we note that the wide MDF and the relatively 
narrow RGB in its CMD (see Fig.~\ref{fig:cmd}) are not easily compatible with one only age 
(or a relatively narrow age range of 1-2 Gyr). 
Stellar isochrones with [Fe/H]$\sim$--1.0 dex and age $\sim$12-13 Gyr should be redder than the RGB of M~54. In order to 
reproduce the position of the metal-rich stars that we observe along the RGB of M~54 we need to assume younger ages, down to some Gyr.  
In particular, the position in the CMD of the most metal-rich stars of the sample, with [Fe/H]$>$--1 dex, is compatible with an age of $\sim$2 Gyr.
A similar result has been found by \citet{liberatori25}, identifying some Sgr field stars (outside the tidal radius of M~54) along the bluest RGB of the galaxy and with [Fe/H]$\sim$--0.5 dex. These stars are compatible with an age of 1-2 Gyr and they can be 
stars formed in an additional burst of star formation of the galaxy or they can be the by-products of mass transfer in binary systems.
In general, the bulk of the MDF of M~54 (with [Fe/H] between --1.8 and --1.0 dex) 
seems to imply an age range between $\sim$13 and $\sim$5 Gyr. Such age range for our sample is consistent with the age-metallicity relation presented by \citet[][see their Fig.~7]{alfaro_cuello_19}. 
However, this age range can be significantly narrowed if we account for a significant He enhancement ($\rm Y\sim0.35$) in the most metal-rich stars.
This is clearly shown in Fig~\ref{fig:cmd_iso}, where three different BaSTI isochrones\footnote{The isochrones were calculated assuming an age of 12 Gyr, a He content of $\rm Y = 0.247$, $\rm [\alpha/Fe] = 0.4$ dex, and three different [Fe/H]: $\rm -1.8, -1.4, -1.1$ dex.} \citep{hidalgo_18} are superimposed to the stars in the CMD. Indeed, by assuming the same age of 12~Gyr for the entire population we can see that the most metal-rich isochrone ($\rm [Fe/H] = -1.1$ dex) is totally outside the locus of the metal-rich component on the CMD. A more appropriate fit is achieved only by assuming a significantly younger age of $\sim4-5$ Gyr (with $\rm Y = 0.247$) or, alternatively, by keeping an age of 12 Gyr but adopting an enhanced He content of $\rm Y \sim 0.35-0.40$.

%--------------------
\subsection{Mg, Al, Si, and K distributions}
We derived the abundances of those elements involved in the complete MgAl burning chain, in particular, Mg, Al, Si, and K abundances for a total of 155, 50, 45, and 75 stars, respectively. 
To derive the Mg elemental abundances, we used the Mg line at 5711 $\AA$, the Mg triplet at 6318-6319 $\AA$, and the line at 8806 $\AA$, according to the used FLAMES setup.
To derive Al abundances we used the doublets at 6696-6698 $\AA$ and at 7835-7836 $\AA$ included in the UVES CD580 and HR18 setups, respectively. 
To derive Si abundances we used about 5-10 lines.
Finally, the K abundances were derived from the second K~I resonance line at 7699 $\AA$\footnote{The first K~I resonance line at 7664 $\AA$ is heavily contaminated by telluric lines, and therefore is not possible to use this line to derive K abundances.} present in the HR18 setup. In this study we corrected K abundances for the Non-LTE effects by interpolating into the grids of \citet{takeda_02}. 

\begin{figure*}[h!]
\centering
\includegraphics[width=\textwidth]{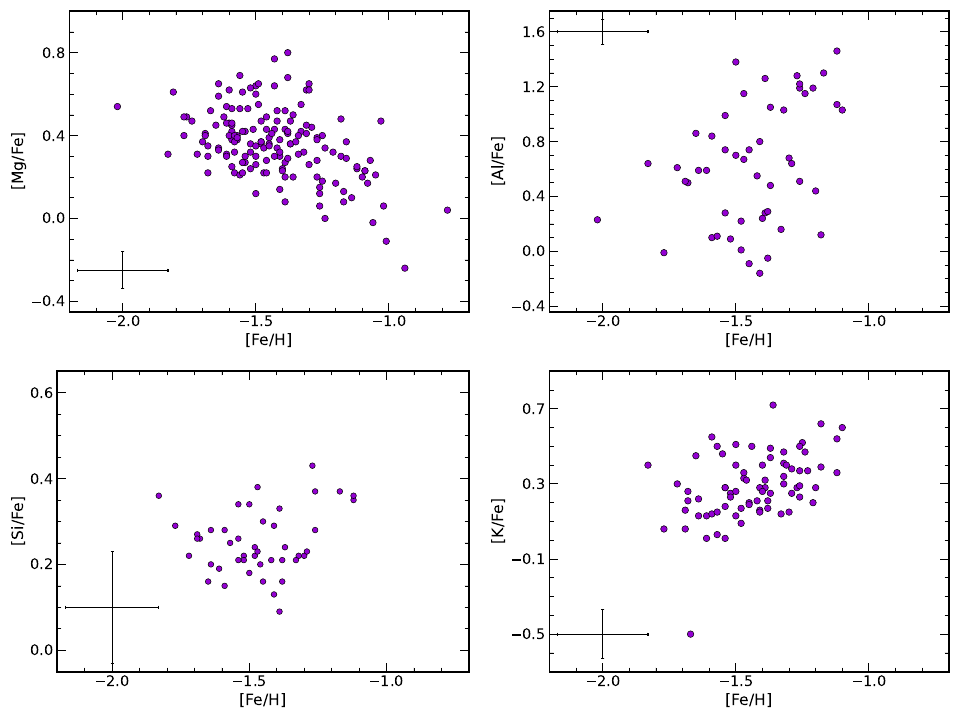}
\caption{The four panels depict the distribution of [Mg/Fe] (top-left), [Al/Fe] (top-right), [Si/Fe] (bottom-left), and [K/Fe] (bottom-right) as function of [Fe/H] for the M~54 stars. The error bar in the corner of each panel represents the typical error associated with the measurements.} \label{fig:mgalsik_fe}
\end{figure*}

Figure~\ref{fig:mgalsik_fe} shows the distribution of [Mg/Fe], [Al/Fe], [Si/Fe], and [K/Fe] abundance ratios as a function of [Fe/H]. 
The stars have enhanced values of [Mg/Fe] ($\sim$+0.4 dex) until [Fe/H]$\sim$--1.2 dex and for higher metallicities an overall decrease down to solar-scaled [Mg/Fe]. The average value of [Mg/Fe] for the stars with [Fe/H]$<$--1.2 dex is comparable with that measured in metal-poor Sgr dSph stars not belonging to M~54 \citep{liberatori25}.
We note a lack of Mg-poor stars in M~54: only three metal-rich stars ($\sim$2\% of the sample with measured [Mg/Fe]) have sub-solar [Mg/Fe] abundance ratios, two of them compatible with solar-value within the uncertainties and one star with [Mg/Fe]$\sim$--0.25 dex. 
\footnote{\citet{carretta_10} identified the target \#3800298 as a Mg-poor star, with [Mg/Fe]=--0.49 dex. 
Our analysis of the same star provides a higher value, [Mg/Fe]=+0.24 dex. For this star only the 6318 \AA\ line is available (at variance with the other stars); because this feature is located on the red wing of a auto-ionization Ca~I line, its abundance is extremely sensitive to the local normalization and the correct modeling of the auto-ionization line.}

For the [Al/Fe] distribution, the behavior is the opposite with respect to the [Mg/Fe] distribution. 
The bulk of the sample exhibits a large star-to-star scatter in [Al/Fe] (from solar value up to $\sim$+1.3 dex), 
while the stars with [Fe/H]$>$--1.2 dex have systematically higher [Al/Fe]. The M~54 stars with lower [Al/Fe] well match 
the [Al/Fe] abundances of the Sgr stars of similar [Fe/H] \citep{liberatori25}, suggesting a similar chemical evolution 
path, while the stars with higher [Al/Fe] can be considered as SG stars within a GC framework.
A small increase with [Fe/H] seems to be visible also for [Si/Fe] but a significantly lower extension. 
Most of the stars have [Si/Fe]$\sim$+0.2 dex, while the few metal-rich stars for which we can measure Si abundances 
have higher [Si/Fe]. 
Finally, also [K/Fe] shows an increasing run with [Fe/H].

%--------------------

The run of the [Mg/Fe], [Al/Fe], [Si/Fe], and [K/Fe] as a function of metallicity presented here resembles the behavior of both [O/Fe] and [Na/Fe] as a function of [Fe/H] presented by \citet{carretta_10} (see their Fig.~15) for the M~54 stars. Indeed, from their analysis emerges that the most metal-poor stars in their sample have both low and high content of [O/Fe] and [Na/Fe], while the most metal-rich stars  are characterized mainly by a high content of [Na/Fe] and low content of [O/Fe].
This seems to be the case also for the [Mg/Fe], [Al/Fe], [Si/Fe], and [K/Fe] distributions in our sample, where our metal-rich component seems to be dominated by the presence of stars with lower (higher) content of Mg (Al, Si, and K) with respect to the metal-poor component. 

%--------------------
\subsection{Mg-Al, Mg-Si, Mg-K anticorrelations}
\begin{figure}[h]
\centering
\includegraphics[width=9 cm]{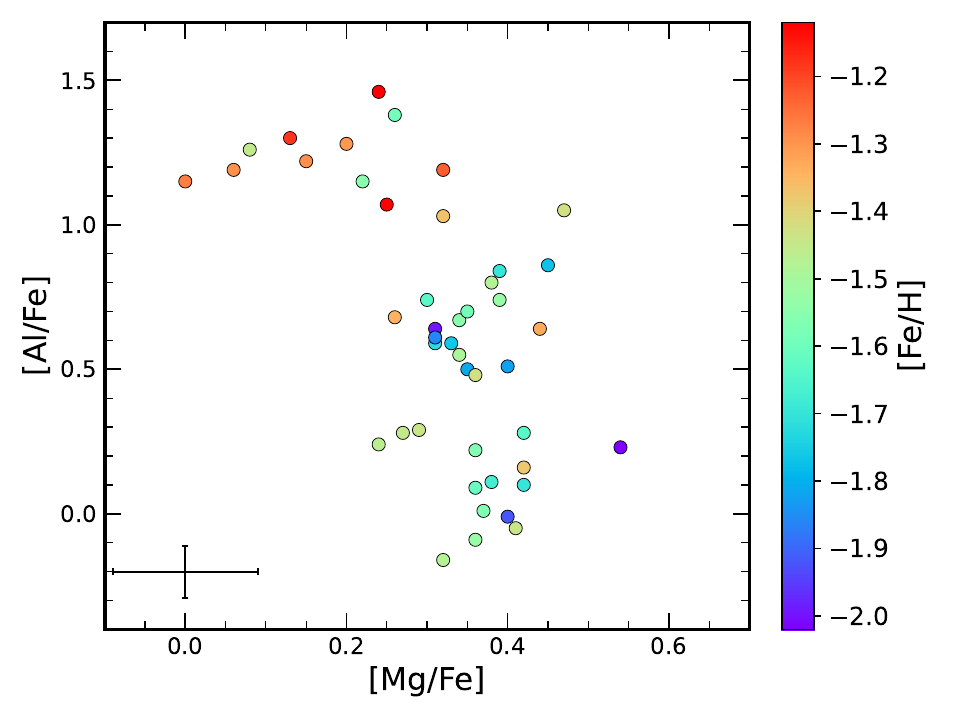}
\caption{Mg-Al anticorrelation for the stars of M~54. The stars are color-coded according to their metallicity. The color scale is shown on the right side. The error bar in the bottom-left corner represents the typical errors.}
\label{fig:mgal}
\end{figure}

We measured both Mg and Al abundances for a total of 45 stars of the sample. 
These stars draw a clear Mg-Al anticorrelation (see Fig.~\ref{fig:mgal}).  
Previously, \citet{carretta_10} found hint of Mg-Al anticorrelation from the analysis of six stars, while  \citet{fernandez_21} detected this anticorrelation in a sample of 19 stars observed within the APOGEE survey \citep{majewski_17}. 
The stars with [Fe/H]$<$--1.4 dex have similar [Mg/Fe] over a large range of [Al/Fe], while 
the most metal-rich stars have [Al/Fe]$>$+1 dex and lower [Mg/Fe].

%--------------------
%--------------------

\begin{figure}[h]
\centering
\includegraphics[width=9 cm]{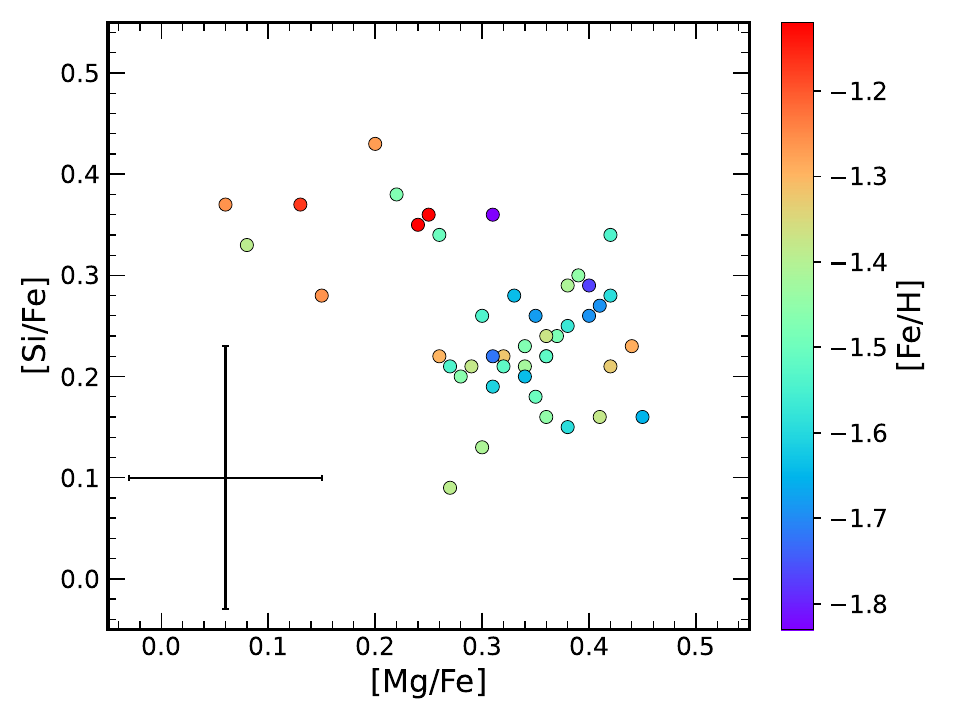}
\caption{Same as Fig.~\ref{fig:mgal}, but for the Mg-Si anticorrelation.}
\label{fig:mgsi}
\end{figure}

%--------------------
Mg and Si abundances were simultaneously available for a total of 45 stars. All the stars cover a similar range of [Mg/Fe] and [Si/Fe], ranging from $\sim 0.05$ to $\sim 0.45$ dex. In Fig.~\ref{fig:mgsi} we can see the overall Mg-Si anticorrelation among the stars of M~54. 
A Spearman correlation test provides a correlation coefficient $C_s$=-0.29 and a p-value of 0.049, enforcing the existence of a real anticorrelation between 
the two abundance ratios, despite the large uncertainties ($\sim$0.15 dex) associated to [Si/Fe]. 
Similar to Mg-Al, also in this case the most metal-rich stars are those with low  [Mg/Fe] and high [Si/Fe]. 
From Fig.~\ref{fig:mgal} and Fig.~\ref{fig:mgsi} we can observe that the stars with the higher content of Al are the same with the higher content of Si, and vice versa.

 %--------------------
%--------------------

\begin{figure}[h!]
\centering
\includegraphics[width=9 cm]{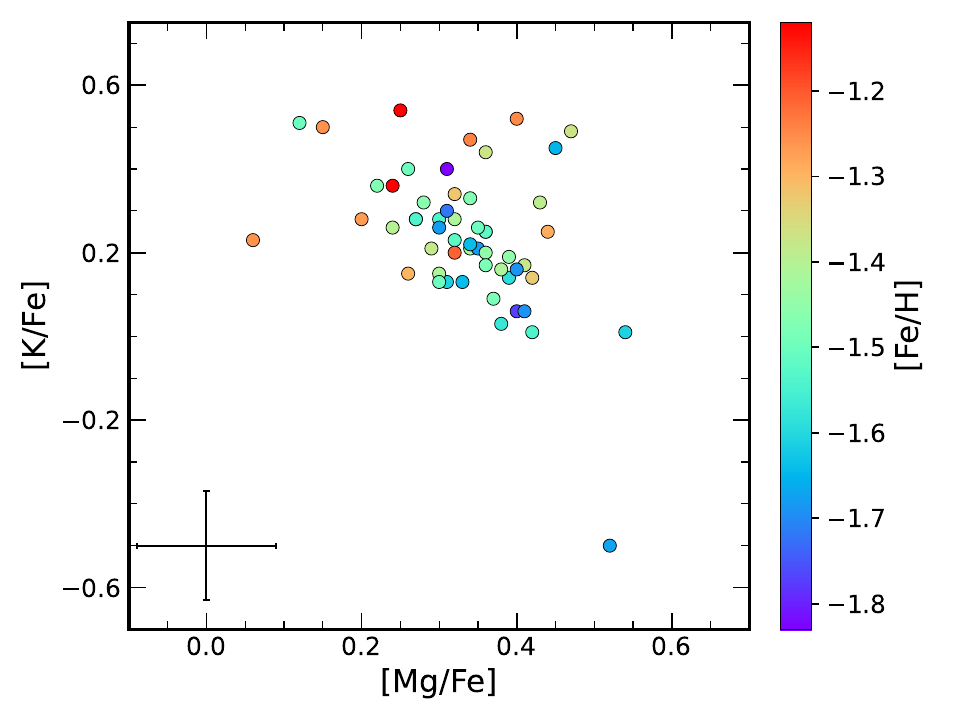}
\caption{Same as Fig.~\ref{fig:mgal}, but for the Mg-K anticorrelation.}
\label{fig:mg_k}
\end{figure}

%--------------------
We measured Mg and K abundances simultaneously for a total of 55 stars.  Figure~\ref{fig:mg_k} shows [K/Fe] as a function of [Mg/Fe].
Indeed, the Spearman correlation test gives $C_s = -0.37$ and a p-value $p = 5.3\times10^{-3}$, confirming the presence of an anticorrelation 
between the two abundance ratios. 
%--------------------
%--------------------

\begin{figure}[h]
\centering
\includegraphics[width=9 cm]{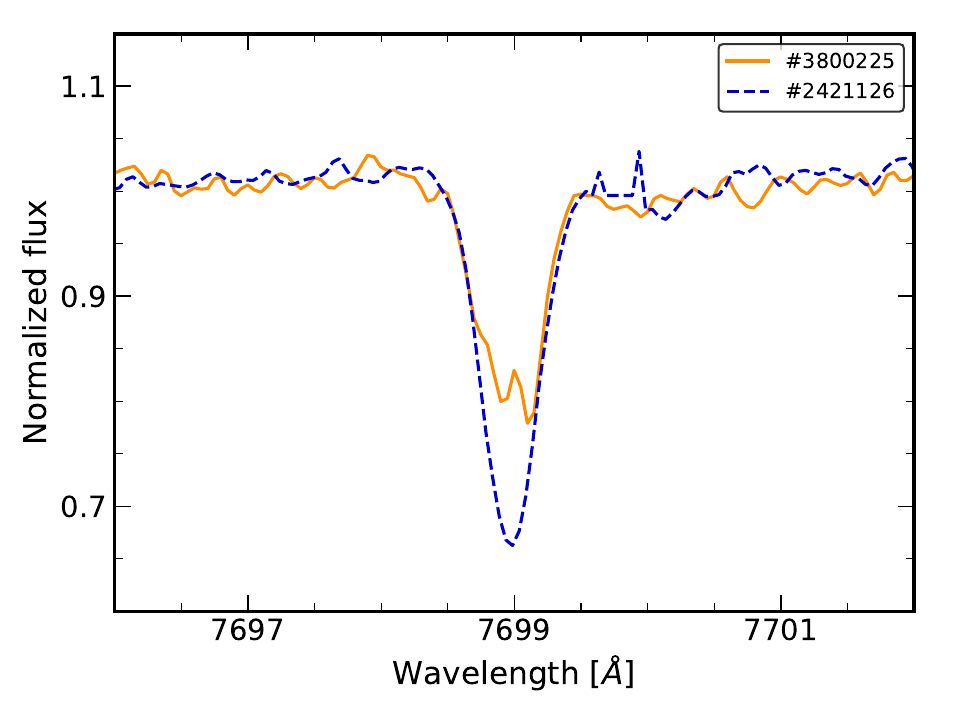}
\caption{Comparison between the HR18 GIRAFFE spectra of \#3800225 (solid orange line) and \#2421126 (dashed blue line) stars around the K line at 7699 $\AA$. The depletion of K of \#3800225 compared to \#2421126 is clearly visible from the comparison.}
\label{fig:overplot_line_K}
\end{figure}

Among the metal-poor group we found one star, named \#3800225, having [K/Fe] $= -0.50$ dex, a value that is much lower than all the other stars making the metal-poor component. This star has [Mg/Fe]=+0.52 dex. We tested whether the derived abundance for this star is real or affected by some artifact introduced during the analysis, but we were not able to find any correlation with the adopted stellar parameters. In Fig.~\ref{fig:overplot_line_K} we can see the K line for the K-poor star compared with a reference star, named \#2421126, having similar parameters ($\rm T_{eff}$, $\rm \log g$, $\rm v_t$, and [Fe/H]) but with [K/Fe]=+0.13 dex, which is compatible with the abundances of the bulk of the metal-poor component. We can see an obvious difference in the line depths, that implies a real difference in the derived K abundances. Therefore, we conclude that the derived [K/Fe] for \#3800225 is real and not an artifact of the analysis. 
%--------------------
%--------------------

The presence of a Mg-K anticorrelation in M~54 was claimed by \citet{carretta_22} that analyzed 42 stars included in our sample. 
Our analysis confirms their result but with greater statistical significance. The Spearman correlation test for the entire dataset analyzed by \citet{carretta_22} provides a $p$-value of 0.05, while the exclusion of the star \#3800298 (the only Mg-poor stars in his analysis but with a normal [Mg/Fe] in our study) leads to a $p$-value of 0.08. 
Our dataset provides a correlation coefficient $C_s = -0.41$ and a $p$-value of $1.99\times10^{-3}$, while, when the anomalous K-poor star is excluded, the dataset provides a $C_s = -0.37$ and a $p$-value of $5.27\times10^{-3}$, confirming the existence of the Mg-K anticorrelation. 
The difference of the statistical significance between the two studies (even if based on the same spectra) is mainly due to the adopted $\rm v_t$, that significantly impacts on the derived K abundances because of the strength of the K lines. 

%------------------------------------------
\section{Discussion} \label{discussion}
The bluest RGB visible in the CMD of the nuclear region of Sgr is dominated 
by the GC-like system M~54, with a minor component corresponding to the metal-poor Sgr stars formed in-situ 
(according to the MDF by \citet{minelli23}, $\sim$14\% of the Sgr stars have [Fe/H]$<$--1.0 dex).
The presence of a [Fe/H] spread confirms that the system was able to retain the ejecta from core-collapse SNe, as occurs in complex stellar systems such as galaxies. On the other hand, the presence of anticorrelations forces us to include, in the interpretative framework of M~54, the chemical-enrichment phenomena typically observed in GCs. 
The existence of a clear Mg-Al anticorrelation in M~54, with stars displaying large variations in the Al abundances, is the signature of Al production through the MgAl chain at very high temperatures \citep{ventura_16,dellagli_18}. The Al variations observed cover $\sim 1.5$ dex, which is quite large. However, such value is not surprising, since large variations in the Al abundances are usually found in those GCs very massive and/or metal-poor \citep{shetrone_96,meszaros_15,pancino_17}. The presence of the Al-Si correlation is the result of a leakage from the MgAl chain into $^{28}\mathrm{Si}$ through proton-capture reaction at very high temperatures \citep{yong_05,meszaros_15,masseron_19}. 
In the following, we provide a possible interpretative scheme for M~54 able to explain the chemical evidences bridging GCs and NSCs.

%--------------------
\subsection{Tentative explanation for the MPs in M~54}
Our interpretative scenario for the formation of MPs in M~54 follows the one proposed by \citet{carretta_10} to explain their results for the Na-O anticorrelation.
Indeed,  both in \citet{carretta_10} and in this work, the metal-rich component is dominated by SG stars with low content of O and Mg, and high content of Na, Al, Si and K, 
while the metal-poor component hosts stars with both low and high abundances of these elements. These findings make the anticorrelations more extended at higher metallicities, 
indicating a higher level of processing by proton-capture reactions in this metallicity regime. 

In this scenario the series of events that led to the formation of the different groups of stars in M~54 started with the formation of the most metal-poor stellar component, followed by the formation of the metal-rich stars, which occurred with a $\sim 10-30$ Myr delay \citep{carretta_10}. These are the FG stars of M~54.  
In this case, while polluter stars of $\rm 6-8~M_{\odot}$ of the metal-poor component evolve into their AGB phase, the massive stars of the metal-rich component continue to explode as core-collapse SNe, preventing the formation of any cooling flow. After the rate of SN explosions becomes low enough, a quiet phase then follows, lasting some tens Myr. In this phase, cooling flow formation towards the central regions is possible for both the metal-poor component (with the contribution also of less massive AGB stars of $\rm 4-6~M_{\odot}$) and the metal-rich one (with AGB stars of $\rm 6-8~M_{\odot}$, the only polluters that had time to evolve into their AGB phase at this time). 

Then, the SG stars form nearly simultaneously in the metal-poor and metal-rich cooling flows. This process lasted until the SN explosion of stars formed in these cooling flows or the onset of type Ia SNe inhibited further star formation. 
This chain of events is enough to explain the fact that the metal-rich regime is composed by stars with a higher level of Mg depletion and Al, Si, and K enhancement. Indeed, the SG stars of the metal-rich regime formed from the gas of the most massive AGB stars (but without significant Mg depletion as explained in Sect.~\ref{sect:abs_mg_poor}), at odds with the SG of the metal-poor component that formed from gas of both massive and less massive AGB stars. 
Moreover, the SG stars of the most metal-rich population should be enhanced in He (up to $\rm Y\sim0.35$) in order to pacify the observed narrow RGB with the broad MDF.

%--------------------
\subsection{The absence of Mg-poor stars in M~54} \label{sect:abs_mg_poor}
In M~54 there is no Mg-poor ([Mg/Fe] $< 0.0$ dex) subpopulation, 
at odds with other systems such as NGC~2419, NGC~2808, NGC~5824 or $\omega$ Centauri \citep{cohen12,mucciarelli_12,mucciarelli_15,mucciarelli_18,alvarez_24}. 
Apart from $\omega$ Centauri, M~54 is the most massive GC. 
This is an indication that even though the chemical composition of SG stars of M~54 shows up the signature of MgAl processing, the degree of the proton-capture nucleosynthesis activated within the polluter stars was not sufficiently high to allow the $\sim 1$ dex depletion of Mg observed in $\omega$ Centauri \citep{alvarez_24}. 

In the theoretical framework of the AGB and super-AGB stars, sufficiently advanced 
proton-capture nucleosynthesis, able to trigger significant depletion of the original Mg, occurs in $\rm 4-8~M_{\odot}$ stars of metallicities of the order of $\rm [Fe/H] \sim -2$ dex or less. In higher metallicity environments such an advanced nucleosynthesis is expected only in AGB stars of mass $6-8$ M$_{\odot}$. In the latter case, in order to form Mg-poor stars in GCs is required that the formation of SG stars took place directly from the gas released  by $\rm M \geq 6~M_{\odot}$ stars into the intracluster medium, within a time interval of the order of $\sim 100$ Myr. 
If the formation of the SG extends over a longer period the interstellar medium would be subsequently enriched by gas from $\rm 4-5~M_{\odot}$ AGB stars that evolve more slowly and whose nucleosynthesis experienced at the base of the envelope, while sufficiently advanced to favor a factor ten or more increase in the Al content, is not extreme enough to destroy significant amounts of Mg \citep{dellagli_18}. Therefore, the formation of SG stars with very low Mg content is easier in metal-poor environments, provided that the SG is formed directly from the AGB ejecta, with no dilution with pristine gas, which has the same chemical composition of FG stars.

On the basis of these arguments, the lack of stars with very low [Mg/Fe] values ([Mg/Fe] $<0.0$ dex) in M~54 is not surprising. While the iron distribution of $\omega$ Centauri exhibits a primary peak in the metal-poor domain, at [Fe/H] $\sim -1.8$ dex \citep{johnson_10,alvarez_24}, the results shown in Fig.~\ref{fig:rvfe} outline the absence of an extended tail to the metal-poor regime in M~54: this indicates that the formation of SG stars with very low Mg was much easier in $\omega$ Centauri than in M~54. In the latter cluster Mg-poor stars could form only in the case that  (1)~rapid star formation occurred,  (2)~directly from the gas released by massive AGB stars of metallicity [Fe/H] $\sim -1.9$ dex, which evidently was not the case.

For the Al the things are slightly different, because the temperatures required for the Al synthesis, of the order of $50$~MK, are less extreme than those required for the Mg destruction, which are slightly below 100 MK \citep{dellagli_18}. This is because the Al content of stars is $\sim 50$ times smaller than Mg, thus even proton-capture reactions by the less abundant $\rm ^{25}Mg$ and $\rm ^{26}Mg$ isotopes of Mg, whose activation requires temperatures significantly cooler than the more abundant $\rm ^{24}Mg$, are sufficient to favor an increase in the Al content of 1 dex or more. This means that a significant increase of Al (e.g., by $\sim 1$ dex in [Al/Fe]) can occur even at moderate low metallicities ([Fe/H]$\sim -1.3/-1.2$ dex), without a corresponding large decrease in [Mg/Fe].

Finally, the modest increase in [Si/Fe] and [K/Fe] can be explained by the fact that in M~54 the Mg is not processed at levels that allow for a significant Si and K production through the secondary chain of the MgAl cycle, at odds with other systems such as NGC~2419 or $\omega$ Centauri: both channels demand temperatures in excess of 100 MK, which are reached only by metal-poor AGBs.
Obviously, this scenario requires an appropriate timing for the gas expulsion from the FG stars and the subsequent formation of the SG. 

\subsection{M~54: a complex cluster}
In this work we are focusing on the dominant old and metal-poor population of the nuclear star cluster lying at the center of the Sgr dSph galaxy. For historical reasons we refer to this component as M~54, however, it is known from the literature as well as from the results presented here that this component is different from the younger and more metal-rich population of the Sgr nucleus both from a kinematic and from a chemical abundance point of view \citep{bellazzini_08, carretta_10,Carretta_10lett,alfaro_cuello_19}. Within this framework, 
in the previous sections, we showed once again that it displays
the chemical patterns typical of MPs observed in GCs and a metallicity spread that is instead typical of galaxies \citep{willman_12}.  
The similarity with $\omega$ Centauri was noted long ago \citep{bellazzini_08, Carretta_10lett}. Indeed $\omega$ Centauri is generally believed to be the nuclear remnant of a disrupted dwarf galaxy \citep[see, e.g.,][and references therein]{bekki03,bekki_19}.

In this context, we can interpret M~54 following the scenario proposed by \citet{alvarez_24} for $\omega$ Centauri, that is the merging of two or more GCs plus some contamination from the in-situ nuclear population, an idea already put forward by \citet{alfaro_cuello_19} for M~54.
In this view M~54 should have formed by the merging  of two GCs with not very different mean metallicity. Indeed, the MDF of M~54 is asymmetric (see Fig.~\ref{fig:rvfe}) but it cannot be resolved  into two distinct components with the extant data.  
The most metal-poor progenitor GC is the dominant component of M~54 and it behaves as a normal GC with intermediate properties, a modest Mg depletion and a significant enhancement of Al. 
The second, less dominant and more metal-rich GC has a larger Mg depletion and a stronger Al enhancement. 
The two original GCs could had different ages, in order to reconcile the narrow RGB of M~54 with the broad MDF, or similar ages if we assume an important He enhancement in the second cluster, although this latter option seems unlikely.

As said, a similar hypothesis has been suggested by \citet{alfaro_cuello_19} using low-resolution MUSE data for the nuclear region of Sgr. 
They found a broad MDF ($\sigma$=0.24$\pm$0.01 dex) for what they called old/metal-poor component of Sgr (and corresponding to M~54 in our nomenclature) and a spread in age of about 1 Gyr, similar to what we argued according to high-resolution spectroscopy and Gaia photometry.

From the work by \citet{bekki_16} emerged that the presence of so called "anomalous" GCs, i.e. those massive systems displaying both MPs and large spreads in metallicity, can be explained by the merging of GCs.  They performed numerical simulations and several tests and found that the merger between GCs more massive than $\rm 3\times10^5~M_{\odot}$ are inevitable within a host dwarf galaxy with a mass in the range between $\rm 3\times10^9~M_{\odot}$ and $\rm 3\times10^{10}~M_{\odot}$. The dark halo of the progenitor of the Sgr dSph is believed to lie in this mass range \citep{niederste_12,deboer_14,gibbons_17}, thus providing a favorable environment for the GCs merging scenario. The infall of these massive clusters is driven by dynamical friction on a scale of a few Gyr. The consequent orbital decay may also favor the merging by bringing both clusters close to the center or they may both reach the center at different epochs and merge there \citep[see also][]{gavagnin}. 
In summary, the merging scenario could explain (1)~the high mass of M~54, (2)~its wide MDF, and (3)~its possible age spread.  
A possible observational test  for this scenario would be to look for a bimodal distribution in [Fe/H]. For a given sample, the power of this test would  be defined by the precision of individual [Fe/H] measures, that should be significantly better than the 0.1-0.2 dex achieved in this work.

%-------------------------------------------------
\section{Summary and conclusions} \label{summary}
We present the largest dataset (233 stars) of high-resolution spectra for RGB stars belonging to the bluest RGB of the nuclear 
region of Sgr dSph and usually associated to the GC-like system M~54. The main results are summarized as follows:
\begin{itemize}
\item We confirmed the wide MDF of M~54, with the bulk of the stars having [Fe/H] between --1.8 and --1.0 dex and the presence of a few stars with [Fe/H]$>$--1.0 dex. An MCMC approach provides [Fe/H]=--1.41$\pm$0.01 dex and an intrinsic scatter $\rm \sigma_{int} = 0.13\pm0.02$ dex for the complete sample, 
and [Fe/H]=--1.43$\pm$0.01 dex and $\rm \sigma_{int}=0.09\pm0.02$ dex considering only the stars with $\rm -2.0<[Fe/H]<-1.0$ dex.
The MDF appears asymmetric but, with the current data, we are not able to resolve possible multiple peaks. 
\item The MDF and the broadening of the RGB in the CMD are not compatible with a single, old age. 
A large age range, from $\sim$13 to $\sim$1-2 Gyr, is necessary to reproduce the RGB width, or alternatively a small age range 
but with a significant enhancement of He (at least for the most metal-rich stars).
\item In M~54 the complete pattern of anticorrelations linked to the MgAl burning chain is evident. In particular, the most metal-rich component displays a higher level of processing by proton-capture reactions. Indeed, the metal-rich component of M~54 has systematically lower [Mg/Fe], and higher [Al/Fe], [Si/Fe], and [K/Fe] compared to the metal-poor component. 
\item We identified a single K-poor star in M 54, \#3800225, exhibiting [K/Fe] $=-0.50$ dex, a value significantly lower than the bulk distribution for stars with similar parameters. As the star displays no other apparent chemical anomalies, and the current dataset (HR18 and HR21 spectra) is insufficient for a detailed characterization, we postpone a comprehensive analysis of this peculiar object to future work.
\item The properties of M~54 cannot be explained neither with a GC-like scenario nor with a galactic chemical evolution. The observed chemical patterns are compatible with a scenario where M~54 formed through the merging of two GCs (not resolved in the MDF due to the uncertainties of individual stars), the metal-poor one with standard characteristics and the more metal-rich one with more pronounced chemical anomalies, a possibly younger than the first GC.

\end{itemize}

In conclusion, M~54 emerges to be a very complex system, where possibly different mechanisms contributed to its chemical enrichment history. 

%-------------------------------------------------
\section{Data availability} 
Full Table~\ref{tab:2} is only available in electronic form at the CDS via anonymous ftp to \url{cdsarc.u-strasbg.fr (130.79.128.5)} or via \url{http://cdsweb.u-strasbg.fr/cgi-bin/qcat?J/A+A/}.

\begin{acknowledgements}
Funded by the European Union (ERC-2022-AdG, "StarDance: the non-canonical evolution of stars in clusters", Grant Agreement 101093572, PI: E. Pancino). Views and opinions expressed are however those of the author(s) only and do not necessarily reflect those of the European Union or the European Research Council. Neither the European Union nor the granting authority can be held responsible for them.
A.M. and M.B. acknowledge support from the project "LEGO – Reconstructing the building blocks of the Galaxy by chemical tagging" (P.I. A. Mucciarelli) granted by the Italian MUR through contract PRIN 2022LLP8TK\_001.
\end{acknowledgements}
\bibliographystyle{aa} % style aa.bst
\bibliography{biblio} % your references Yourfile.bib

\end{document}